\documentclass{article}
\usepackage{nicefrac}
\usepackage{amsfonts}
\usepackage{amssymb}
\usepackage{amsmath,mathrsfs,dsfont}
\usepackage{graphicx}
\usepackage{appendix}
\usepackage{cite}
\usepackage{color}
\begin{document}

\title{UV/IR mixing as a twisted Poincar\'{e} anomaly.}
\author{A. Pinzul\thanks{apinzul@unb.br} \\
\\
\emph{Universidade de Bras\'{\i}lia}\\
\emph{Instituto de F\'{\i}sica}\\
\emph{70910-900, Bras\'{\i}lia, DF, Brasil}\\
\emph{and}\\
\emph{{International Center of Condensed Matter Physics} }\\
\emph{C.P. 04667, Brasilia, DF, Brazil} \\
}
\date{}
\maketitle

\begin{abstract}
We analyze symmetries of the 1-loop effective action of $\phi^4$ noncommutative field theory. It is shown, that despite the
twisted Poincar\'{e} invariance of the classical noncommutative action, its 1-loop quantum counterpart lacks this invariance. Though Noether analysis of the model is somewhat obscure, it is still possible to interpret this symmetry breaking as a quantum anomaly due to inappropriate choice of the quantization method.
\end{abstract}

\newpage

\section{Introduction}

Based on a quite general argument \cite{Doplicher:1994tu}, it is believed that one or another form of a noncommutative (NC) space-time should emerge as a quasi-classical approximation to quantum gravity in a sense that quantum fields on such a space-time will probe the quantum gravity scale through noncommutativity. This observation, as well as the appearance of NC field theory in super strings \cite{Seiberg:1999vs}, sparked considerable interest in noncommutative physics. Most of the efforts have been devoted to the analysis of theories on a NC space-time with constant noncommutativity, a so-called Moyal space. (Literature on the subject is vast, for a review see, e.g. \cite{Douglas:2001ba}) This is the simplest possible type of noncommutativity, defined by the following commutation relation between the coordinates
\begin{equation}
[x^\mu , x^\nu] = i \theta^{\mu\nu} \label{cr}\ .
\end{equation}
Here $\theta^{\mu\nu}$ gives the scale of quantum gravity, usually it is thought to be related to the Planck scale. It is clear that if $\theta^{\mu\nu}$ is some non-dynamical constant, which is the case in string theory, then (\ref{cr}) breaks Lorentz invariance.

The study of Lorentz-breaking theories has a long history by now \cite{Colladay:1998fq}. Up to the first order in $\theta^{\mu\nu}$, NC quantum field theory with constant noncommutativity is just a special case of the general situation (e.g., see discussion in \cite{Chatillon:2006rn}). The quantization of NC field theories (NCFT) had been performed within this framework, i.e quantization of a Lorentz violating theory. Perturbative analysis led to some noncommutative corrections but also, what is more important, to a quite universal property of NCFT -- UV/IR mixing \cite{Minwalla:1999px}. (Though, see some recent proposals on models without UV/IR mixing \cite{Grosse:2003nw,Grosse:2004yu,Gurau:2008vd}.) This property, which is due to the non-local nature of noncommutativity, put an end to hopes that noncommutativity (at least in the simple form of a constant $\theta^{\mu\nu}$) would serve as a UV regulator, which was the major motivation for one of the first works on a NC space-time \cite{Snyder:1946qz}.

After works \cite{Wess:2003da,Chaichian:2004za}, another possibility to look at NCFT had become possible: instead of thinking in terms of broken Lorentz invariance, one could now think about NCFT as a theory with preserved, but {\it twisted} Lorentz symmetry.\footnote{Though we analyze the full Poincar\'{e} invariance, only the Lorentz part is non-trivially twisted, see Section \ref{tnc}.} On the classical level, this is just a point of view that does not add anything new to the picture. But the transition to quantization changes the situation drastically: one has to maintain this new twisted symmetry on quantum level. Essentially, there exist two approaches to quantization of NCFT with twisted Lorentz symmetry:
\begin{itemize}
\item The quantization should be done according to the usual rules. In other words, the usual Feynman rules, derived from the path integral with the standard quantum measure should be used to calculate quantum processes \cite{Chaichian:2008ge}. In this approach, all the technical results and conclusions about NCFT, including UV/IR mixing, remain untouched.\cite{Chaichian:1998kp,Tureanu:2006pb,Zahn:2006wt}
\item The quantization scheme should be drastically modified via deformation of canonical commutation relations (CCR) for creation/annihilation operators \cite{Pinzul:2005gx,Balachandran:2005eb} (see also \cite{Oeckl:2000eg,Sasai:2007me,Lukierski:2011mj}). Implementation of this approach has resulted in quite dramatic consequences: absence of UV/IR mixing in pure matter theories \cite{Balachandran:2005pn}, non-trivial possibilities for gauge theories \cite{Balachandran:2007kv} and many others.\cite{Balachandran:2007vx,Balachandran:2006ib,Balachandran:2008gr,Akofor:2007fv,Akofor:2008ae,Balachandran:2009sq,Balachandran:2011pd,Leineker:2010uw}
\end{itemize}
Both approaches claim to preserve the twisted Lorentz symmetry on quantum level. The arguments from both sides are based on the analysis of Green's functions, construction of the Hilbert space (carrying the correct representation of the twisted symmetry) as well as other considerations. Despite all the efforts, a consensus is still absent.

In this paper, we address the problem of quantization of theories with the twisted Lorentz symmetry from a different point of view. We choose one of the two quantization schemes, namely the standard one (with undeformed CCR), and analyze the symmetries of a quantum object -- 1-loop effective action. This approach is, in some sense, close to the one used in analyzing quantum anomalies. The best case scenario would be to present an analog of a classically  conserved Noether charge, corresponding to the twisted Lorentz symmetry, and analyze quantum corrections to the conservation law. Unfortunately, as we will see, Noether analysis is still quite obscure in this case and we will have to directly look for symmetries of the quantum action. The result of this analysis appears to be unambiguous: 1-loop effective action does not respect the twisted Lorentz invariance! Moreover, it is possible to show that this invariance is broken by exactly those diagrams that lead to the appearance of the UV/IR mixing -- hence the title of the paper.

The plan of the paper is following. In Section \ref{tnc}, we analyze the difficulties of the Noether analysis in NCFT with the twisted symmetry. In Section \ref{1lea}, we give a sketchy review of the construction of the 1-loop effective action following \cite{Minwalla:1999px}. Section \ref{symmetry} deals with the analysis of the symmetries of the 1-loop effective action. In Conclusion and discussion section, we discuss obtained result as well as indicate possible directions for the future development. In particular, we argue that the reason for the quantum anomaly is the intrinsic incompatibility of the chosen quantization scheme with the geometry of the noncommutative space-time. The majority of technical details about the Noether analysis is collected in Appendix. The whole analysis is done on the example of an interacting scalar field.

\section{On twisted Noether charges}\label{tnc}

The Noether analysis of noncommutative theories is not an easy task. This is due to the fact, that instead of the usual Lie-type Lorentz symmetry, noncommutative theories respect its twisted version \cite{Wess:2003da,Chaichian:2004za}. There are at least two different approaches to the problem. In the first \cite{Aschieri:2008zv}, the Moyal-type noncommutativity is dynamical. This leads to a new scalar degree of freedom and a natural restoration of the Lorentz invariance, which makes possible the usual analysis of conserved Noether currents. In the second \cite{AmelinoCamelia:2007wk}, one uses noncommutative parameters in the twisted Poincar\'{e} transformations, which allows to preserve the usual Leibnitz rule; this, in turn, allows to extract Noether currents (though, explicitly it has been done only for the case of a free scalar field). Both, approaches, from our point of view, have some disadvantages. In the first one, we don't deal with the actual (twisted) symmetry of a noncommutative theory. The price to pay is the presence of an extra field. In the second, along with a not very clear physical reason for noncommutative parameters, one ends up with the impossibility to realize pure boosts (though this may not be a disadvantage but a mere physical reality).

Here we will not perform a general analysis of the problem but rather consider a specific noncommutative model of an interacting scalar field on a Moyal space-time. The goal of this section is to demonstrate difficulties of the Noether analysis as well as to argue that having the twisted symmetry does not in general guarantee existence of the corresponding conserved charges. This should help to appreciate the fact that one does not need conserved charges to analyze symmetries at the quantum level, see Section \ref{symmetry}. In passing, we will obtain a closed expression for the energy-momentum tensor, $T_{\mu\nu}$, which has correct transformation properties under the twisted symmetry.\footnote{The existence of a conserved $T_{\mu\nu}$ is expected, because (\ref{cr}) is formally translation invariant. Below we discuss the question of uniqueness of the energy-momentum tensor.} The results of this section, especially on $T_{\mu\nu}$, will be analyzed and applied elsewhere.

Our approach is based on the direct analysis of the following model:
\begin{equation}
S = \int d^4 x {\cal L}_* \ , \mbox{where } {\cal L}_* = \frac{1}{2}\partial_\mu \phi * \partial^\mu \phi + \frac{1}{2}m^2\phi * \phi + \frac{\lambda}{4!} \phi*\phi*\phi*\phi \ .
\label{action}
\end{equation}
Here $*$ is the usual Moyal star-product:
\begin{equation}
(f*g)(x) := f(x) e^{\frac{i}{2}\theta^{\mu\nu}\overset{\leftarrow}{\partial}_\mu \overset{\rightarrow}{\partial}_\nu} g(x) \ .
\label{star}
\end{equation}
This action leads to the following equation of motion
\begin{equation}
-\Box\phi + m^2 \phi + \frac{\lambda}{3!}\phi*\phi*\phi = 0 \ .
\label{eom}
\end{equation}
It is well known that the action (\ref{action}) explicitly violates the Poincar\'{e} invariance. This is not unexpected, because (\ref{cr}) is clearly not Lorentz invariant (though translational invariance is not broken) in the case of nonzero and non-dynamical $\theta^{\mu\nu}$, i.e. when $\theta^{\mu\nu}$ does not transform as a tensor. But now, the usual Lie-type symmetry is not natural anymore. This is due to the fact that the noncommutative product (\ref{star}) does not agree with the usual co-product, which leads to the usual Lie-type transformations of composite fields:
\begin{eqnarray}\label{problem}
m_* \circ\Delta (h) (\phi \otimes \psi ) \ne h \triangleright m_*  (\phi \otimes \psi ) \ ,
\end{eqnarray}
where $m_*  (\phi \otimes \psi ) = \phi * \psi$ is the noncommutative product (\ref{star}), $h$ is an arbitrary element of the symmetry algebra and $\Delta (h)=h\otimes\mathds{1}+\mathds{1}\otimes h$ is the standard (untwisted) co-product. This suggests that a new type of invariance, the twisted Poincar\'{e} symmetry, should be considered \cite{Chaichian:2004za}. We will not go into a detailed discussion of the twisted invariance and provide here only the formulas we will need for the further consideration, namely the expressions for the twisted co-product, which cures the problem in (\ref{problem}) making it into an equality:
\begin{eqnarray}
& & \Delta_\theta (P_\mu) = \mathds{1} \otimes P_\mu +  P_\mu \otimes \mathds{1} \ , \nonumber\\
& & \Delta_\theta (M_{\mu\nu}) = \mathds{1} \otimes M_{\mu\nu} +  M_{\mu\nu} \otimes \mathds{1} + S_{\mu\nu}^{\sigma\gamma} (P_\gamma\otimes P_\sigma ) \ ,
\label{coproduct}
\end{eqnarray}
where
$$
S_{\mu\nu}^{\sigma\gamma} = -\frac{1}{2}\left( \theta_{[\mu}^{\ \ \sigma}\delta_{\nu ]}^{\ \ \gamma} + \theta^{\gamma}_{\ [\mu}\delta_{\nu ]}^{\ \ \sigma}\right)
$$
and $P_\mu$ and $M_{\mu\nu}$ are the standard Poincar\'{e} generators:
\begin{equation}
P_\mu = i\partial_\mu \mbox{\ \ \ and\ \ \ } M_{\mu\nu} = ix_{[\mu}\partial_{\nu ]} \ .
\label{generators}
\end{equation}
Let us denote the most general transformation by $h$:
\begin{equation}
h := \epsilon^\mu P_\mu + \omega^{\mu\nu} M_{\mu\nu} \ .
\label{generaltrans}
\end{equation}
This will generate the usual variation of the space-time coordinates
\begin{equation}
\delta x^\mu = -ih[x^\mu] \equiv \epsilon^\mu +\omega^{\rho\nu}(x_{\rho}\delta_{\nu}^{\ \mu} - x_{\nu}\delta_{\rho}^{\ \mu}) \ .
\label{coord}
\end{equation}
and, using the fact that $\phi (x)$ is a scalar field, i.e. $\phi' (x')=\phi (x)$, the variation of the field
\begin{equation}
\delta \phi (x) = ih[\phi (x)] \equiv - (\partial_\mu \phi)\delta x^\mu \ .
\label{field}
\end{equation}
It is very important that in (\ref{field}) there is no star-product between $\partial_\mu \phi$ and $\delta x^\mu $ as it follows from the fact that, even in the twisted case, the Poincar\'{e} symmetry acts in the usual way on a single field (or a single-particle sector).

Now, using (\ref{coproduct}), we can define the action of the twisted Poincar\'{e} symmetry on a star-product of any two fields, $\psi$ and $\chi$ (either a scalar filed $\phi$ or its derivatives)
\begin{equation}
\delta(\psi * \chi) = \delta (\psi) * \chi + \psi * \delta (\chi) + i\omega^{\mu\nu}S_{\mu\nu}^{\sigma\gamma} (\partial_\sigma \psi * \partial_\gamma \chi ) \ .
\label{leibniz}
\end{equation}
In particular, in the case of a pure translation, we have the usual Leibnitz rule, while in general this rule gets deformed by the last term in (\ref{leibniz}). Actually, the deformed Leibnitz rule is all we need to calculate the action of the twisted Poincar\'{e} symmetry on any composite object defined on the NC space-time. In particular, this observation immediately leads to the conclusion that only composite objects, like the Lagrangian (\ref{action}), that are made of covariant fields with star-products will transform covariantly, see  more on this in Section \ref{symmetry}.

Let us apply the formulae above to find a variation of the action (\ref{action})
\begin{eqnarray}\label{varaction}
\delta S = \int d^4 x (\bar\delta{\cal L}_* + {\cal L}_*\partial_\mu \delta x^\mu) \ .
\end{eqnarray}
Here $\bar\delta{\cal L}_*$ is a complete variation both due to the variation of a field (\ref{field}) and coordinates (\ref{coord})
\begin{equation}
\bar\delta{\cal L}_* = \delta{\cal L}_* + (\partial_\mu {\cal L}_*)\delta x^\mu \ ,
\label{varlagr}
\end{equation}
where now $\delta{\cal L}_*$ is the variation due to fields only. As in the commutative case, the variation of the measure (the second term in (\ref{varaction})) and the variation of the Lagrangian due to coordinates combine in a total derivative
\begin{equation}
\delta S = \int d^4 x \left(\delta{\cal L}_* + \partial_\mu ({\cal L}_*\delta x^\mu )\right) \ .
\label{varlagr1}
\end{equation}
If we could, using the equation of motion (\ref{eom}), show that the first term in (\ref{varlagr1}) also can be represented as a total derivative, we would immediately arrive at the noncommutative analog of Noether's theorem. Before we proceed, two (trivial) comments are in order: 1) While in deriving the equation of motion (\ref{eom}) one is allowed (and has to) integrate by parts dropping surface terms, in obtaining Noether's currents the only thing that is permitted is the use of the equations of motion. This is, of course, because in deriving conserved currents we use arbitrary integration region; 2) Because of this, one does not have a cyclic property under integral, which explicitly uses integration by parts, and, as a result, one has to carefully take care of the order of terms in products of fields while calculating the variation (\ref{varaction}) (see Appendix).

To keep the presentation simple and not to distract from the main results, we put all the calculations of $\delta{\cal L}_*$ as well as all the notations in Appendix, the result being, see Eq.(\ref{Lvarcomplete})
\begin{eqnarray}
\delta{\cal L}_*=
\frac{1}{2}\partial_\mu\left( \delta \phi * \partial^\mu \phi + \partial^\mu \phi * \delta \phi + \frac{i}{2}\omega^{\alpha\beta}S_{\alpha\beta}^{\sigma\gamma} [\partial_\sigma \partial^\mu \phi , \partial_\gamma  \phi ]_* \right)-\nonumber \\ - \frac{\lambda}{4!}\partial_\mu\left( \frac{\sinh\Delta}{\Delta}([\delta\phi ,\phi]_*, X^\mu(\phi *\phi ))  - i\omega^{\alpha\beta}S_{\alpha\beta}^{\sigma\gamma}\frac{\cosh\Delta - 1}{\Delta}\left( \partial_\sigma \phi * \phi , X^\mu(\phi * \partial_\gamma \phi \right))\right)+\nonumber\\
+ i\frac{\lambda}{4!}\omega^{\alpha\beta}S_{\alpha\beta}^{\sigma\gamma}(\partial_\sigma \phi * \phi )( \phi * \partial_\gamma \phi ) \ .
\label{Lvariation}
\end{eqnarray}
We see that while the first two lines in (\ref{Lvariation}) are total derivatives, the third one is not. This kills the usual argument in the passage from the variation of an action to the corresponding conserved current. Of course, this does not mean that (at the classical level) the theory does not respect the twisted invariance: the total variation (\ref{varaction}) is still perfectly zero. What this means, that now it is not clear whether we have a conserved quantity associated to this symmetry or not.

It is important to notice that the problem with the conserved current exists only for the Lorentz part of the twisted Poincar\'{e} symmetry: the problematic term in (\ref{Lvariation}) is proportional to $\omega^{\mu\nu}$ -- the Lorentz parameters. In the case of the pure translations, when $\omega^{\mu\nu} = 0$, we do have a conserved current. It is nothing but a twisted generalization of the energy-momentum tensor, $T^{\mu\nu}$. Using (\ref{coord}), (\ref{field}), (\ref{varlagr1}) and (\ref{Lvariation}), we obtain
\begin{eqnarray}\label{emt}
T^{\mu\nu} = -\frac{1}{2}\partial^\mu \phi * \partial^\nu \phi -\frac{1}{2}\partial^\nu \phi * \partial^\mu \phi - \frac{\lambda}{2} \frac{\sinh\Delta}{\Delta}\left( [\phi,\partial^\nu \phi]_* , X^\mu (\phi*\phi) \right) + \eta^{\mu\nu} \mathcal{L}_* \ .
\end{eqnarray}
The fact that $T^{\mu\nu}$ is obtained from the full twisted invariance guarantees that its conservation is frame independent. Indeed, though  $T^{\mu\nu}$ itself is not written just in terms of the star-product, its divergence, $\partial_\mu T^{\mu\nu}$, involves only the product (\ref{star}). This means that the conservation law $\partial_\mu T^{\mu\nu} = 0$ is consistent with the twisted Poincar\'{e} symmetry, i.e. it is twisted covariant.

Even in the commutative case, $T^{\mu\nu}$ is defined up to a divergence of an arbitrary antisymmetric tensor of the third order. In the case of noncommutativity this freedom is much larger. This is due to the existence of a new object, $\theta^{\mu\nu}$, which can be used to contract indices coming from extra derivatives. Let us illustrate this on the example of a free scalar field where all the essential points are already present.

In this case, the action is given by (\ref{action}) without the interaction term and can be written in two equivalent forms:
\begin{eqnarray}
S = \int d^4 x \frac{1}{2}(\partial_\mu \phi * \partial^\mu \phi + m^2\phi * \phi ) \equiv \int d^4 x \frac{1}{2}(\partial_\mu \phi \partial^\mu \phi + m^2\phi^2 )\ ,
\end{eqnarray}
where we used the possibility to remove one Moyal star by integration by parts. It is obvious that this action respects both, the twisted and the usual Poincar\'{e} symmetries. Then using the twisted symmetry of the first form, we have from (\ref{emt})
\begin{eqnarray}\label{emtfreetwist}
T^{\mu\nu}_{\theta} = -\frac{1}{2}\partial^\mu \phi * \partial^\nu \phi -\frac{1}{2}\partial^\nu \phi * \partial^\mu \phi + \frac{1}{2}\eta^{\mu\nu} (\partial_\rho \phi * \partial^\rho \phi + m^2\phi * \phi ) \ .
\end{eqnarray}
On the other hand, from the untwisted invariance of the second form of the action we have
\begin{eqnarray}\label{emtfree}
T^{\mu\nu}_0 = -\partial^\mu \phi \partial^\nu \phi + \frac{1}{2}\eta^{\mu\nu} (\partial_\rho \phi \partial^\rho \phi + m^2\phi^2 ) \ ,
\end{eqnarray}
i.e. the usual commutative energy-momentum tensor. It is easily checked that both tensors are conserved on the equation of motion:
\begin{eqnarray}
\partial_\mu T^{\mu\nu}_{\theta} = \partial_\mu T^{\mu\nu}_{0} = 0 \ .
\end{eqnarray}

We can write $T^{\mu\nu}_{\theta}$ in terms of $T^{\mu\nu}_{0}$ plus higher order derivative terms:
\begin{eqnarray}\label{TT}
T^{\mu\nu}_{\theta} = T^{\mu\nu}_{0} - \frac{i}{2}\partial_\rho \frac{\cosh\Delta - 1}{\Delta}\left\{ (\partial^\mu \phi , X^\rho \partial^\nu \phi) - \eta^{\mu\nu}[(\partial_\sigma \phi , X^\rho \partial^\sigma \phi ) + m^2 (\phi , X^\rho \phi )] \right\}\ .
\end{eqnarray}
The second term in (\ref{TT}) explicitly breaks the usual Poincar\'{e} invariance, while being separately conserved.\footnote{It is also important to note, that in the case of the pure space-space noncommutativity, i.e. when $\theta^{0i}=0$, the second term in (\ref{TT}) does not introduce higher order {\it time} derivatives.} So, even in the free case, the choice between $T^{\mu\nu}_{\theta}$ and $T^{\mu\nu}_{0}$ should be dictated by the underlying symmetry of the space-time. We are planning to pursue this line elsewhere.

\section{1-loop effective action}\label{1lea}

To quantize our model (\ref{action}), we have to choose between the possibilities discussed in Introduction. To answer the main question of this paper: whether the usual quantization scheme respects the twisted invariance, we will use the standard quantization based on the standard, i.e. undeformed, CCR (see the discussion in the concluding section), which leads to the usual Feynman rules. In other words, we will think of our noncommutative model as a special type of non-local theory, defined on (commutative) Minkowski space. Specifically, we are interested in 1-loop quadratic effective action, $S^{(2)}$. This calculation was done in \cite{Minwalla:1999px}. Here we just repeat main steps leading to the result.

As it was demonstrated in \cite{Minwalla:1999px}, in spite of the same quantization procedure, there is one major difference between commutative and noncommutative models: planar and non-planar diagrams contribute differently. This is due to the fact that while vertices in planar diagrams look the same as in the commutative case and, as a result, the contribution is exactly the same as in the commutative case (modulo some overall phase factor depending on the external momenta), in the non-planar case vertices are really non-local, see below, leading to the drastic deviation from the commutative result. In our case, the relevant diagrams are shown on Fig.1.

\begin{figure}[htb]
\begin{center}
\leavevmode
\includegraphics[scale=0.5]{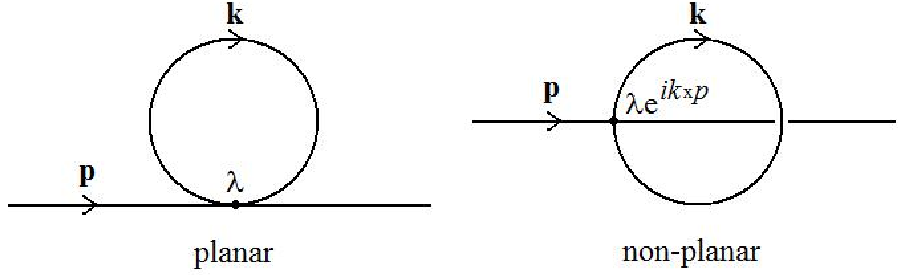}
\end{center}
\caption{The diagrams that contribute to $\Gamma^{(2)}_{1PIplanar}$ and $\Gamma^{(2)}_{1PInon-planar}$.}
\label{fig:scaling}
\end{figure}

As we said, the contribution of the planar diagram to the 1PI two-point function is exactly as in the commutative case:
\begin{equation}\label{planar}
\Gamma^{(2)}_{1PIplanar}=\frac{\lambda}{3(2\pi)^4}\int\frac{d^4 k}{k^2+m^2} \ .
\end{equation}
On the other hand, a vertex of the non-planar diagram will contain a non-local factor $e^{ik\times p}$, where $k\times p := k_\mu \theta^{\mu\nu}p_\nu$, leading to the following contribution (the relative factor of 2 between the two expressions is due to the fact that there are twice as many planar diagrams):
\begin{equation}\label{nonplanar}
\Gamma^{(2)}_{1PInon-planar}=\frac{\lambda}{6(2\pi)^4}\int\frac{d^4 k}{k^2+m^2}e^{ik\times p} \ .
\end{equation}
Introducing UV cutoff $\Lambda$, integrals (\ref{planar}),(\ref{nonplanar}) can be easily calculated by the standard methods leading to the result
\begin{eqnarray}
\Gamma^{(2)}_{1PIplanar}= \frac{\lambda}{48\pi^2}\left( \Lambda^2 - m^2\ln\left( \frac{\Lambda^2}{m^2}\right)+\mathcal{O}(1)\right) \ , \label{planarfin}\\
\Gamma^{(2)}_{1PInon-planar}= \frac{\lambda}{96\pi^2}\left( \Lambda^2_{nc} - m^2\ln\left( \frac{\Lambda^2_{nc}}{m^2}\right)+\mathcal{O}(1)\right) \ . \label{nonplanarfin}
\end{eqnarray}
Here $\Lambda_{nc}$ is the crucial modification of the cutoff due to noncommutativity:
\begin{eqnarray}\label{lambdanc}
\Lambda_{nc} = \frac{\Lambda}{\sqrt{1+\Lambda^2 p_{nc}^2}} \ ,
\end{eqnarray}
where $p_{nc}^\mu:=p_\nu \theta^{\nu\mu}$. Aside from this modification, both contributions look exactly the same (modulo the aforementioned factor 2). But while $\Gamma^{(2)}_{1PIplanar}$ leads to the usual mass renormalization, $\Gamma^{(2)}_{1PInon-planar}$ results in the famous UV/IR mixing. This can be seen from the behavior of $\Lambda_{nc}$: considering first $p_{nc}\rightarrow 0$, i.e. IR limit, we recover the usual UV divergence, $\Lambda_{nc}\rightarrow \Lambda$; if we instead first take UV limit, i.e. $\Lambda\rightarrow\infty$, then $\Lambda_{nc}\rightarrow\frac{1}{\sqrt{p_{nc}^2}}$, which produces divergences in IR regime $p_{nc}\rightarrow 0$.

Now we will write down the main object of our interest -- the quadratic part of the 1-loop effective action, which is trivially obtained from (\ref{planarfin}) and (\ref{nonplanarfin})
\begin{eqnarray}\label{S2}
S^{(2)}=\frac{1}{2}\int d^4 p\ \phi(p)\phi(-p)\left( p^2 + m^2 + \frac{\lambda}{48\pi^2}\left( \Lambda^2 - m^2\ln\left( \frac{\Lambda^2}{m^2}\right)\right) + \frac{\lambda}{96\pi^2}\left( \Lambda^2_{nc} - m^2\ln\left( \frac{\Lambda^2_{nc}}{m^2}\right) \right) \right) \ .
\end{eqnarray}
As we said, $\Gamma^{(2)}_{1PIplanar}$ just renormalizes the mass, $m_{ren}^2 = m^2 + \frac{\lambda}{48\pi^2}\left( \Lambda^2 - m^2\ln\left( \frac{\Lambda^2}{m^2}\right)\right)$, while $\Gamma^{(2)}_{1PInon-planar}$ leads to a non-trivial {\it momentum dependent} contribution, which cannot be absorbed into a redefinition of parameters of the model. In the next section, we will analyze this term from the point of view of the twisted symmetry.

\section{Symmetry of the 1-loop effective action}\label{symmetry}

Due to the modified cut-off (\ref{lambdanc}), the quadratic part of the one-loop effective action (\ref{S2}) is a highly non-local object. To establish its transformation properties under the twisted Poincar\'{e} symmetry (\ref{coproduct}), we first notice that the transformation law can be established for any bi-linear expression of the form
\begin{equation}\label{bilinear}
D(\partial_x ,\partial_y)\phi(x)\psi(y)|_{x=y} \ ,
\end{equation}
where $D(\partial_x ,\partial_y)$ is some pseudo-bidifferential operator. As we have already commented after Eq.(\ref{leibniz}), the whole effect of the twisting for the composite fields is captured by the deformed Leibnitz rule (\ref{leibniz}). In other words, the transformation of (\ref{bilinear}) under the general element $h$ (\ref{generaltrans}) is given by
\begin{eqnarray}
D(\partial_x ,\partial_y)\phi(x)\psi(y)|_{x=y} \xrightarrow{\text{\it\ h\ }} D(\partial_x ,\partial_y)\left(\Delta_\theta (h)\phi(x)\otimes\psi(y)\right)|_{x=y} = \nonumber\\
= D(\partial_x ,\partial_y) \left[\delta\phi(x)\psi(y)+\phi(x)\delta\psi(y)+i\omega^{\mu\nu}S^{\sigma\gamma}_{\mu\nu}(\partial_\sigma\phi(x)\partial_\gamma\psi(y))\right]_{x=y} \ . \label{bilineartrans}
\end{eqnarray}
In particular, if we take
\begin{equation}
D(\partial_x ,\partial_y)=e^{\frac{i}{2}\theta^{\mu\nu}\partial_x\partial_y} \ ,\nonumber
\end{equation}
we will obtain the correct transformation of $(\phi*\psi) (x)$ given by (\ref{leibniz}). From this it should be clear that only for this form of $D(\partial_x ,\partial_y)$ we will have covariant objects, e.g. $(\phi*\psi) (x)$ will be a scalar. (See the comment after (\ref{leibniz}).) On the other hand, taking $D = \mathds{1} \otimes \mathds{1}$, we immediately see that $(\phi\psi)(x)\equiv\phi(x)\psi(x)$ is not a scalar under the twisted action (though, it is, of course, a scalar for the usual, \textit{untwisted}, action).

Now, using (\ref{bilineartrans}), we can establish the transformation properties of the effective action (\ref{S2}). It is pretty clear from the expressions for $\Lambda_{nc}$ and $S^{(2)}$, Eqs. (\ref{lambdanc}) and (\ref{S2}), that the non-locality of $S^{(2)}$ cannot be written in terms of a star-product, i.e. in terms of a {\it local} product on {\it NC space}.\footnote{The expression (\ref{S2}) can be treated either as non-locality on the commutative space-time or, again, as non-locality but now on the noncommutative space-time. In both cases, the underlying symmetry is not respected. More on this in the conclusion.} Here, by the locality on NC space we mean that everything can be written in terms of the $*$-product and without extra infinite number of derivatives that cannot be cast into this form. So, we do not expect any nice transformation properties of (\ref{S2}). Indeed, this can be shown by the explicit calculation using (\ref{bilineartrans}).

Let us start by noticing that applying (\ref{bilineartrans}) to $\phi(p)\phi(-p)$ trivially produces
\begin{eqnarray}\label{30}
\delta_\omega (\phi(p)\phi(-p)) = \delta_\omega \phi(p)\phi(-p) + \phi(p)\delta_\omega \phi(-p) \ ,
\end{eqnarray}
i.e. the usual commutative result. Here $\delta_\omega$ is a transformation with the Lorentz parameters $\omega^{\mu\nu}$ defined as in (\ref{generaltrans}), $\delta_\omega := \omega^{\mu\nu}M_{\mu\nu}$ ($M_{\mu\nu}$ is the usual Lorentz generator in momentum representation), and we used antisymmetry of $S^{\rho\sigma}_{\mu\nu}$ defined in (\ref{coproduct}).\footnote{There should not be any illusion that due to (\ref{30}) the twisted action in momentum space is trivial: $\phi(p)\phi(-p)$ does not contain the whole information about star-product -- for this we need the product of the fields at \textit{different} momenta times the phase factor, which vanishes when these momenta are equal. Eq.(\ref{30}) rather corresponds to the well-known fact that we can integrate out one star.} Then we immediately arrive at the conclusion that the first three terms of (\ref{S2}), $\phi(p)\phi(-p)\left( p^2 + m^2 + \frac{\lambda}{48\pi^2}\left( \Lambda^2 - m^2\ln\left( \frac{\Lambda^2}{m^2}\right)\right)\right)$, transform as a scalar, i.e. this part of the effective action is twist invariant. This is, of course, somewhat trivial result and follows from the observation that the usual quadratic action respects both twisted and usual symmetries, see the related discussion at the end of Section \ref{tnc}. Let us now analyze the last term in (\ref{S2}), which comes from the non-planar diagrams. What we are going to show is that this term is not twist invariant, i.e. it does not transform under the twisted action of the Lorentz symmetry (\ref{bilineartrans}) as a scalar. This can be seen by comparing two transformations, the first due to (\ref{bilineartrans}) and the second assuming that this term is a scalar:
\begin{eqnarray}\label{defleibnitz}
\delta_\omega \left( \Lambda^2_{nc} - m^2\ln\left( \frac{\Lambda^2_{nc}}{m^2}\right) \right)(\phi(p)\phi(-p)) \overset{(\ref{bilineartrans})}{=} \left( \Lambda^2_{nc} - m^2\ln\left( \frac{\Lambda^2_{nc}}{m^2}\right) \right) (\delta_\omega \phi(p)\phi(-p) + \phi(p)\delta_\omega \phi(-p)) \ ,
\end{eqnarray}
\begin{eqnarray}\label{scalar}
\delta_\omega \left( \Lambda^2_{nc} - m^2\ln\left( \frac{\Lambda^2_{nc}}{m^2}\right) \right)(\phi(p)\phi(-p)) \overset{scalar}{=} \left( \Lambda^2_{nc} - m^2\ln\left( \frac{\Lambda^2_{nc}}{m^2}\right) \right) (\delta_\omega \phi(p)\phi(-p) + \phi(p)\delta_\omega \phi(-p)) +\nonumber\\
+4\omega^{\mu\nu} p_\mu \theta_{\nu\sigma} p^{\sigma}_{nc} \Lambda^4_{nc}\left( 1-\frac{m^2}{\Lambda^2_{nc}}\right) \phi(p)\phi(-p) \ .
\end{eqnarray}
Comparing (\ref{defleibnitz}) and (\ref{scalar}), we find that the difference is not zero leading to the anomaly of the quantum action with respect to the twisted Poincar\'{e} symmetry, namely its Lorentz part
\begin{eqnarray}\label{anomaly}
\delta_\omega S^{(2)} =  \frac{\lambda}{48\pi^2}\int d^4 p\ \omega^{\mu\nu} p_\mu \theta_{\nu\sigma} p^{\sigma}_{nc} \Lambda^4_{nc}\left( 1-\frac{m^2}{\Lambda^2_{nc}}\right) \phi(p)\phi(-p)\ .
\end{eqnarray}

The straightforward inspection of (\ref{anomaly}) shows that this is not zero (or total derivative). Also it is clear, that this result is entirely due to the non-planar diagrams, which are also the source of the UV/IR mixing. Hence, one can say that the UV/IR mixing is the quantum anomaly of the twisted Poincar\'{e} symmetry.

As we saw in Section \ref{tnc}, it is a quite non-trivial task to associate a conserved quantity to the twisted symmetry (there remains a question whether this quantity even exists). If this conserved current does exist, then (\ref{anomaly}) would naturally lead to anomaly in the corresponding conservation law. On the other hand, the existence of a quantum anomaly is not related to the existence of the classically conserved current. The best way to see this is to use Fujikawa's approach to anomalies \cite{Fujikawa:1979ay,Fujikawa:1980eg}. In this method, one looks at the symmetry of a genuinely quantum object -- the partition function
\begin{eqnarray}\label{partfunc}
Z=\frac{1}{N}\int D\phi\ e^{i S[\phi]} \ .
\end{eqnarray}
This object consists of two fundamentally different parts: a \textit{classical} action $S[\phi]$ and a \textit{quantum} measure $D\phi$. $S[\phi]$ is invariant under some symmetry by the very fact of the existence of this symmetry at the classical level. In the case of the usual symmetry this, of course, leads, by Noether's procedure, to a conserved current. But note, that the existence of this current is not a requirement but, rather, a consequence of the symmetry. To check whether a quantum theory is anomalous or not, it is enough to have some, e.g. twisted, symmetry of the classical action $S[\phi]$. This is because the possible anomaly comes from the quantum measure $D\phi$ in the form of the possibly non-trivial Jocobian (see, e.g. \cite{Fujikawa:2004cx} for the details). In other words, the origin of quantum anomaly is non-invariance of the measure in the path integral. Then we can see that the existence of the non-trivial anomaly (\ref{anomaly}) in the case of a scalar field on a non-commutative space-time tells us that the corresponding quantum measure does not respect the classical symmetry, which, in our case, is the twisted Poincar\'{e} symmetry. But what is the path integral measure in our case? It is exactly the choice of the quantization scheme or the canonical commutation relations. As we stressed at the beginning of Section \ref{1lea}, to obtain the 1-loop effective action (\ref{S2}) we used the usual Feynman rules based on the usual ``commutative'' CCR. All the differences in the form of the non-trivial non-planar diagrams came from the noncommutative \textit{classical} action. This implies that our perturbation theory is the expansion of the path integral with the standard ``commutative'' quantum measure (and the noncommutative action). Then our result on anomaly is a clear signal that this measure is not an appropriate choice if we want to preserve the twisted symmetry at the quantum level. This, in turn, implies that the standard quantization method is inconsistent with the twisted symmetry.\footnote{The somewhat analogous result was obtained in the non-relativistic case and rotational symmetry \cite{Sinha:2011dt}.}

\section{Conclusion and discission}

The main objective of this paper is to analyze the survival of the classical symmetry after quantization. While in the commutative case the answer to the question whether the quantum theory is anomalous or not depends only on the underlying classical symmetry and the field content (and, possibly, the geometry of space-time, e.g. the number of dimensions), in the case of the noncommutative theory we have another source of ambiguity -- the choice of the quantization method. In this paper, we have argued that, in general, the standard ``commutative'' quantization applied to a non-commutative field theory does not preserve the classical twisted Poincar\'{e} invariance. At least for the case of the interacting scalar field on NC space, this has been demonstrated by the direct inspection of the symmetry properties of a quantum object -- the 1-loop effective action. Here we discuss this result from the point of view of the approach used in \cite{Balachandran:2005eb,Balachandran:2005pn}.

In \cite{Balachandran:2005eb,Balachandran:2005pn}, the analysis of the quantization of noncommutative theories with the twisted symmetry on classical level was performed along the lines of canonical quantization. It was shown that to realize the correct representation of the twisted symmetry on the quantum Hilbert space, i.e. on the Fock space, one has to (un)twist quantum field in quite a unique way. This, in turn, immediately leads to the twisted canonical commutation relations (for the details see \cite{Balachandran:2007vx}):
\begin{eqnarray}\label{CCR}
a_{k_2}a^\dag_{k_1}=e^{i\theta_{\mu\nu}k^\mu_1 k^\nu_2}a^\dag_{k_1}a_{k_2} + 2k^0_1\delta^{(3)} (k_1-k_2)\ ,
\end{eqnarray}
where $a_k$ and $a^\dag_k$ are the usual creation and annihilation operators of a scalar field.\footnote{It is worth noting, that this is in complete agreement with the more formal approach within a framework of Braided QFT \cite{Oeckl:2000eg,Sasai:2007me} (see, e.g., Eq.(102) in \cite{Sasai:2007me}).} If we look at this CCR from the path integral point of view, we are forced to conclude that the path integral measure has to be modified in order to produce such commutation relations. This follows from the trivial observation that (\ref{CCR}) is the \textit{quantum} commutation relation and all the quantum information is stored in the path integral measure. So, this is in perfect agreement with the conclusion of the present paper.

Let us now discuss from this point of view our observation that the famous UV/IR mixing could be treated as quantum anomaly of the twisted symmetry. In \cite{Pinzul:2005gx,Balachandran:2005pn}, it was shown that the UV/IR mixing is absent if one uses (\ref{CCR}) to quantize a noncommutative field theory.\footnote{The price to pay is quite high: all the traces of noncommutativity are erased from the scattering processes in any pure matter theory. To bring back the potentially observable effects, one has either to consider a non-trivial coupling of non-commutative matter to gauge fields \cite{Balachandran:2007kv,Balachandran:2006ib,Balachandran:2008gr} or to use more involved physical setups, e.g. \cite{Akofor:2007fv} discusses the modification of the cosmic microwave background in this approach.} This strongly supports our interpretation of the UV/IR mixing as the quantum anomaly: as we have just discussed, the quantization scheme used in \cite{Pinzul:2005gx,Balachandran:2005pn} preserves the twisted symmetry at the quantum level and as a consequence the UV/IR mixing ($=$ quantum anomaly) is absent in this approach.

The discussion above shows that the main result of the present paper should not come as a big surprise. It is still quite important, because it allows us to unambiguously and conclusively prove the incompatibility of the usual quantization method and the twisted Poincar\'{e} invariance. The fundamental reason for this is, from our point view, the incompatibility of this quantization with the geometry of the space-time. By this we mean the following. Choosing to quantize our noncommutative field theory by the ``commutative'' prescription, i.e. using the same quantum measure in the path integral, we automatically choose the commutative space-time with its untwisted Poincar\'{e} symmetry as our geometry and treat noncommutativity just as a special form of non-locality. Then we should not be surprised that the resulting quantum theory is neither Poincar\'{e} invariant (this invariance is broken by the non-locality) nor twisted Poincar\'{e} invariant (this is not the symmetry of the commutative space-time). If we insist that our space-time is genuinely noncommutative, in particular that its symmetry is the twisted Poincar\'{e} invariance, we have to look for a new quantization scheme consistent with this geometry, the approach of \cite{Balachandran:2005eb,Balachandran:2005pn} being one possible candidate. It is instructive to stress that in this case a typical noncommutative Lagrangian, as in (\ref{action}), is \textit{local} from the point of view of this geometry, i.e. depends only on the finite number of derivatives (the $*$-product (\ref{star}) is non-local only from the point of view of the commutative space-time).

There are still many things to do to have a complete picture. We will discuss just two the most obvious.

We have shown that the commutative measure does not seem to be the appropriate choice to quantize theories with the twisted symmetry. So, what is the correct measure? The partial answer is already contained in \cite{Balachandran:2005eb,Balachandran:2005pn} and the related papers, see also \cite{Oeckl:2000eg,Sasai:2007me}. Using the approach of the deformed CCR's, one can, in principle, calculate all the n-point functions, which is equivalent to perturbative definition of the quantum measure. Still, it is quite desirable to construct this measure explicitly and verify following Fujikawa that the twisted anomaly is absent.

The second pressing issue to address is the existence of the twisted Noether charges. We touched this in Section \ref{tnc} with a mostly negative result -- the problem resists the brutal force approach. Still we managed to clarify the importance of the choice of underlying symmetry, which was demonstrated by the construction of the energy-momentum tensor. The existence of two non-equivalent energy-momentum tensors even in the free case could have some far reaching consequences. This line is the subject of our current research.

\section*{Acknowledgement}
The author acknowledges partial support of CNPq under grant no.308911/2009-1.

\section*{Appendix}

Here we provide the details of the derivation  of the result (\ref{Lvariation}). We will separately analyze each of the three terms in (\ref{action}). Before we proceed, we introduce some notation and derive some properties of the star-product (\ref{star}).

Let us introduce a differential operator, $\Delta$ (not to be confused with the co-product (\ref{coproduct}), which we will not use explicitly), acting on a tensor product of fields:
\begin{equation}
\Delta (\psi\otimes\chi) := \frac{i}{2}\theta^{\mu\nu}(\partial_\mu\psi\otimes\partial_\nu\chi) \ .
\end{equation}
Using this operator, we can write the star-product (\ref{star}) as follows
\begin{equation}
\psi *\chi = m\circ e^\Delta (\psi\otimes\chi)\ \ ,\ \mbox{where}\ \ m\circ (\psi\otimes\chi)(x) = \psi (x)\chi (x) \ .
\label{star1}
\end{equation}
Because we will need this type of expressions very often, we will use a short-hand notation for it:
\begin{equation}
m\circ F(\Delta) (\psi\otimes\chi):= F(\Delta)(\psi ,\chi)\ \ ,\ \mbox{where $F(\Delta)$ is any (nice) function of $\Delta$} \ .
\end{equation}
Now several important properties can be easily proven \cite{Aschieri:2008zv}. Everywhere below, $X^\mu := i\theta^{\mu\nu}\partial_\nu$
\begin{eqnarray}
&i)& \psi *\chi = \psi \chi + \Delta\frac{e^\Delta -1}{\Delta}(\psi ,\chi) = \psi \chi + \frac{1}{2}\partial_\mu \frac{e^\Delta -1}{\Delta}(\psi ,X^\mu\chi) \nonumber\\
&ii)& \{\psi ,\chi\}_*:=\psi *\chi + \chi * \psi = 2\psi \chi + 2\Delta\frac{\cosh\Delta -1}{\Delta}(\psi ,\chi) = 2\psi \chi + \partial_\mu \frac{\cosh\Delta -1}{\Delta}(\psi ,X^\mu\chi) \nonumber\\
&iii)& [\psi ,\chi]_*:=\psi *\chi - \chi * \psi = 2\Delta\frac{\sinh\Delta}{\Delta}(\psi ,\chi) = \partial_\mu \frac{\sinh\Delta}{\Delta}(\psi ,X^\mu\chi)\nonumber
\end{eqnarray}
It is important, that the formal division by $\Delta$ is well defined in i)-iii) because all the relevant functions are not singular. Now we are ready to calculate $\delta{\cal L}_*$.
\begin{itemize}
\item[I.] Kinetic term, $\partial_\mu \phi * \partial^\mu \phi$.
\begin{eqnarray}
\delta(\partial_\mu \phi * \partial^\mu \phi) = \delta (\partial_\mu \phi) * \partial^\mu \phi + \partial_\mu \phi * \delta (\partial^\mu \phi) + i\omega^{\alpha\beta}S_{\alpha\beta}^{\sigma\gamma} (\partial_\sigma \partial_\mu \phi * \partial_\gamma \partial^\mu \phi ) = \nonumber\\
\partial_\mu\left( \delta \phi * \partial^\mu \phi + \partial^\mu \phi * \delta \phi + \frac{i}{2}\omega^{\alpha\beta}S_{\alpha\beta}^{\sigma\gamma} [\partial_\sigma \partial^\mu \phi , \partial_\gamma  \phi ]_* \right)
-\delta \phi * \Box \phi - \Box \phi * \delta \phi - \frac{i}{2}\omega^{\alpha\beta}S_{\alpha\beta}^{\sigma\gamma} [\partial_\sigma \Box \phi , \partial_\gamma  \phi ]_*
\end{eqnarray}
\item[II.] Mass term, $\phi * \phi$.
\begin{equation}
\delta(\phi * \phi) = \delta \phi * \phi + \phi * \delta \phi + \frac{i}{2}\omega^{\alpha\beta}S_{\alpha\beta}^{\sigma\gamma} [\partial_\sigma \phi ,\partial_\gamma \phi ]_*
\label{massterm}
\end{equation}
\item[III.] Interaction term, $\phi*\phi*\phi*\phi$.

This term is the most non-trivial one. Here the effects of noncommutativity are the most important.
\begin{eqnarray}
\delta (\phi*\phi*\phi*\phi) = \delta\phi*\phi*\phi*\phi + \phi*\delta\phi*\phi*\phi + \phi*\phi*\delta\phi*\phi + \phi*\phi*\phi*\delta\phi + \nonumber \\
+i\omega^{\alpha\beta}S_{\alpha\beta}^{\sigma\gamma}\left(\partial_\sigma\phi*\partial_\gamma\phi*\phi*\phi + \partial_\sigma\phi*\phi*\partial_\gamma\phi*\phi + \partial_\sigma\phi*\phi*\phi*\partial_\gamma\phi \right.+ \nonumber \\ \left. +\phi*\partial_\sigma\phi*\partial_\gamma\phi*\phi + \phi*\partial_\sigma\phi*\phi*\partial_\gamma\phi + \phi*\phi*\partial_\sigma\phi*\partial_\gamma\phi\right) \ .
\end{eqnarray}
\end{itemize}
Now we will combine I-III and use the equation of motion (\ref{eom}) to obtain $\delta{\cal L}_*$.
\begin{eqnarray}
\delta{\cal L}_*=\delta\left( \frac{1}{2}\partial_\mu \phi * \partial^\mu \phi + \frac{1}{2}m^2\phi * \phi + \frac{\lambda}{4!}\phi*\phi*\phi*\phi  \right) = \nonumber \\
\frac{1}{2}\partial_\mu\left( \delta \phi * \partial^\mu \phi + \partial^\mu \phi * \delta \phi + \frac{i}{2}\omega^{\alpha\beta}S_{\alpha\beta}^{\sigma\gamma} [\partial_\sigma \partial^\mu \phi , \partial_\gamma  \phi ]_* \right)-\nonumber \\ - \frac{\lambda}{4!}\left( [[\delta\phi ,\phi]_*, \phi *\phi]_* \right) + i\frac{\lambda}{4!}\omega^{\alpha\beta}S_{\alpha\beta}^{\sigma\gamma}\{ \partial_\sigma \phi * \phi , \phi * \partial_\gamma \phi \}_* \ .
\label{Lvariation1}
\end{eqnarray}
The first term in (\ref{Lvariation1}) is explicitly a total derivative. Using ii) and iii) properties of the star-product we can show that almost all of the remaining terms are also total derivatives:
\begin{eqnarray}
\delta{\cal L}_*=
\frac{1}{2}\partial_\mu\left( \delta \phi * \partial^\mu \phi + \partial^\mu \phi * \delta \phi + \frac{i}{2}\omega^{\alpha\beta}S_{\alpha\beta}^{\sigma\gamma} [\partial_\sigma \partial^\mu \phi , \partial_\gamma  \phi ]_* \right)-\nonumber \\ - \frac{\lambda}{4!}\partial_\mu\left( \frac{\sinh\Delta}{\Delta}([\delta\phi ,\phi]_*, X^\mu(\phi *\phi ))  - i\omega^{\alpha\beta}S_{\alpha\beta}^{\sigma\gamma}\frac{\cosh\Delta - 1}{\Delta}\left( \partial_\sigma \phi * \phi , X^\mu(\phi * \partial_\gamma \phi \right))\right)+\nonumber\\
+ i\frac{\lambda}{4!}\omega^{\alpha\beta}S_{\alpha\beta}^{\sigma\gamma}(\partial_\sigma \phi * \phi )( \phi * \partial_\gamma \phi ) \ .
\label{Lvarcomplete}
\end{eqnarray}
The last term cannot be brought to a total derivative form. Exactly this term is responsible for the failure of the naive Noether analysis, see the discussion in Section \ref{tnc}.

\end{document}